\documentclass{elsart}
\usepackage{latexsym}
\usepackage{amssymb}
\usepackage[dvips]{graphicx}

\newcommand{\derr}[2]{\frac{d #1}{d #2}}
\newcommand{\beq}[0]{\begin{equation}}
\newcommand{\eeq}[0]{\end{equation}}

\newcommand{\mc}[1]{\mathcal{#1}}

\begin{document}
\newlength{\caheight}
\setlength{\caheight}{12pt}
\multiply\caheight by 7
\newlength{\secondpar}
\setlength{\secondpar}{\hsize}
\divide\secondpar by 3
\newlength{\firstpar}
\setlength{\firstpar}{\secondpar}
\multiply\firstpar by 2

\begin{frontmatter}
\rightline{FNT/T-2001/19}
\vskip 48pt
\title{A Path Integral Way to Option Pricing}
\author[1,2]{Guido Montagna}\author[2,1]{ and Oreste Nicrosini}
\address[1]{Dipartimento di Fisica Nucleare e Teorica, Universit\`a di Pavia\\
Via A. Bassi 6, 27100, Pavia, Italy}
\address[2]{Istituto Nazionale di Fisica Nucleare, sezione di Pavia\\
Via A. Bassi 6, 27100, Pavia, Italy}
\author[3]{Nicola Moreni}
\address[3]{Universit\'e Pierre et Marie Curie Paris 6, Place Jussieu \\
75252 Paris Cedex 05, France}
\begin{abstract}
An efficient computational algorithm to price 
financial derivatives is presented. It is based on a path integral 
formulation of the pricing problem. It is shown how the path integral 
approach can be worked out in order to obtain fast and accurate 
predictions for the value of a large class of 
options, including those with path-dependent and early exercise
features. As examples, the application of the method to European and 
American options in the Black-Scholes model is illustrated. 
A particularly simple and fast semi-analytical approximation for the price 
of American options is derived.
The results of the algorithm are compared with those obtained with the standard 
procedures known in the literature and found to be in good agreement.\\
\end{abstract}

\end{frontmatter}
\newpage
\section{Introduction}
\label{intro}
The classical theory of option pricing is based on the results
found in 1973 by Black and Scholes~\cite{BS} and, independently,
Merton~\cite{Me}. Their pioneering work starts from the 
basic assumption that the asset prices follow 
the dynamics of a particular stochastic process 
(geometric Brownian motion), so that they have a lognormal
distribution~\cite{Hu,PB}. In the case of an efficient market with no arbitrage
possibilities, no dividends and constant volatilities, they found 
that the price of each financial derivative is
ruled by an ordinary partial differential equation, 
known as the Black-Scholes-Merton (BSM) formula. In the most simple
case of a so-called European option, the BSM equation can
be explicitly solved to obtain an analytical formula 
for the price of the option~\cite{Hu,PB}. When we consider other financial derivatives, 
which are commonly traded in real markets and allow  
anticipated exercise and/or depend on the history of the 
underlying asset, the BSM formula fails to give an analytical result. 
Appropriate numerical procedures have been developed in the 
literature to price exotic financial derivatives with path-dependent features, 
as discussed in detail in Refs.~\cite{Hu,WDH,bpa}. The aim of this work is to 
provide a contribution to the problem of efficient option pricing in 
financial analysis, showing how  it is possible to use path integral methods to 
develop a fast and precise algorithm for the evaluation of 
option prices.

The path integral method, which traces back to the original work of 
Wiener and Kac in stochastic calculus~\cite{W,K} and of Feynman in quantum 
mechanics~\cite{FH}, 
is today widely employed in chemistry and physics, 
and very recently in finance too~\cite{Rt,Rt2,B,I,M}, because it gives the possibility of 
applying powerful analytical and numerical techniques~\cite{S}. Following recent studies 
on the application of the path integral approach to the 
financial market as appeared in the econophysics literature (see 
Refs.~\cite{B,M} for a comprehensive list of references), this paper is 
devoted to present an original, efficient path integral algorithm to price financial 
derivatives, including those with path-dependent and early exercise features, and 
to compare the results with those obtained with the standard procedures known 
in the literature. 

The paper is organized as follows.
In Section \ref{option} the basic ideas of the classical theory of 
option pricing are summarized, discussing the computational complexity
associated to the evaluation of the 
price of a path-dependent option and reviewing the 
standard numerical procedures adopted in the literature. 
In Section \ref{path} the path integral approach to option pricing 
is described and analytically developed in order to obtain 
an efficient procedure 
for the calculation of the transition probability associated 
to a given stochastic model of asset evolution. 
Theoretical and computational details to obtain 
fast predictions for path-dependent options are also described. 
As applications of the method, numerical
results for European and American options in the BSM model are
given in Section \ref{numerical}, together with comparisons with 
results known in the literature. A 
particularly simple and very quick semi-analytical approximation 
for the price of an American option is derived in 
Section \ref{american}, by exploiting 
the possibility of 
anticipated exercise for any time before 
the expiration date.  Conclusions and possible perspectives are 
drawn in Section \ref{conclusion}.

\section{Option pricing: theory and numerical procedures}
\label{option}

\subsection{Classical theory and path-dependent options}
\label{optiont}
The basic ingredient for the development of a theory
of option pricing  is a suitable model for the time
evolution of the asset prices.
The assumption of the BSM model is that the price $S$ of an
asset is driven by a geometric Brownian motion 
and verifies the stochastic differential equation (SDE) \cite{Hu,PB}
 \beq
 dS=\mu S dt+ \sigma S dw ,
 \eeq
 which, by means of the It\^o lemma, can be cast in the form of 
 an arithmetic Brownian motion for the logarithm of $S$
 \beq
 d(\ln S)=A dt+ \sigma dw ,
 \label{eq:ito}
 \eeq
where $\sigma$ is the volatility, $A\doteq\left(\mu-\sigma^{2}/2\right)$, 
$\mu$ is the drift parameter and
$w$ is the realization of a Wiener process. By virtue of the 
properties of a Wiener process, eq.~(\ref{eq:ito}) may be written as
 \beq
 d(\ln S)=A dt+ \sigma\epsilon\sqrt{dt} ,
 \label{eq:itow}
 \eeq
where $\epsilon$ follows from a standardized normal distribution with mean 0 and variance
1.\\
Thus, in terms of the logarithms of the asset prices 
$z'\doteq\ln S',z\doteq\ln S$,
the conditional transition probability $p(z'|z)$ to have at the
time $t'$ a price $S'$ under the hypothesis that the price was $S$ at the time $t<t'$ is
given by~\cite{PB,Rt}\footnote{The correct way to indicate conditional transition probabilities is
$p(z',t'|z,t)$. We omit the times in order to simplify the notation.}

 \beq
 p(z'|z)=\frac{1}{\sqrt{2\pi (t'-t) \sigma^{2}}}
 \exp{\left\{-\frac{[z'-(z+A(t'-t))]^{2}}{2\sigma^{2}
 (t'-t)}\right\}} ,
 \label{eq:gauss}
 \eeq
which is a gaussian distribution with mean $z+A(t'-t)$ and variance $\sigma^{2}(t'-t)$.\\
If we require the options to be exercised
only at specific times $t_{i},i=1,\cdots,n$, the asset price, between two
consequent times $t_{i-1}$ and $t_{i}$, will follow eq.~(\ref{eq:itow}) and the
related transition probability will be
 \beq
 p(z_{i}|z_{i-1})=\frac{1}{\sqrt{2\pi \Delta t
 \sigma^{2}}}\exp{\left\{-\frac{[z_{i}-(z_{i-1}+A\Delta t)]^{2}}{2\sigma^{2}
 \Delta t}\right\}} ,
 \label{eq:pdf}
 \eeq
with $\Delta t = t_{i} - t_{i-1}$.

A time-evolution model for the asset price is strictly necessary 
in a theory of option pricing because
 the fair price at time $t=0$ of an option
$\mc{O}$, without possibility of anticipated exercise before the
expiration date or maturity $T$ (a so-called European option), 
is given  by the scaled expectation value~\cite{Hu}
 \beq
 \mc{O}(0)=e^{-r T}E[\mc{O}(T)] ,
 \eeq
 where $r$ is the risk-free interest and
 $E[\cdot]$ indicates the mean value, which can be computed only if 
 a model for the asset underlying the option is understood. 
 For example, the value $\mc{O}$ of an European call
option at the maturity $T$ will be 
$\max\{S_{T}-X,0\}$, where $X$ is the strike price, while 
for an European put option the value $\mc{O}$ at the maturity will 
be $\max\{X-S_{T},0\}$. It is worth emphasizing, for what follows, that the 
case of an European option is particularly simple, since in such a situation 
the price of the option can be evaluated by means of analytical formulae,
which are obtained by solving the BSM partial differential equation 
with the appropriate boundary 
conditions~\cite{Hu,PB}. On the other hand, many further 
kinds of options are present in the financial markets, such as American options 
(options which can be exercised at any time up to the expiration date) and 
exotic options~\cite{Hu}, i.e. derivatives with complicated payoffs or whose value 
depend on the whole time evolution of the underlying asset and not just on its value 
at the end. For such options with path-dependent and early exercise features 
no exact solutions are available and pricing them correctly is a great challenge.
 \\
Actually, in the case of options with possibility
of anticipated exercise before the expiration date, the above discussion 
needs to be generalized, by introducing a slicing 
of the time interval $T$. Let us consider, for definiteness, 
the case of  an option which can be exercised within the maturity
but only at
the times $t_{1}=\Delta t, t_{2}=2\Delta t,\ldots,t_{n}=n\Delta t\doteq T.$
At each time slice $t_{i-1}$ the value $\mc{O}_{i-1}$ of the
option will be the maximum between its expectation value at the time
$t_{i}$ scaled with $e^{-r\Delta t}$ and its value in the case of
anticipated exercise
$\mc{O}_{i-1}^{Y}$~\footnote{\footnotesize 
For example, the value $\mc{O}$ of a call
option in the case of exercise at the time $t_{i-1}$ will be 
$\max\{S_{i-1}-X,0\}$, $X$ being the strike price.}. If $S_{i-1}$ 
denotes the price of the underlying
asset at the time $t_{i-1}$, we can thus write for each
$i=1,\ldots,n$
 \beq
 \mc{O}_{i-1}(S_{i-1})=\max\left\{\mc{O}_{i-1}^{Y}(S_{i-1}),e^{-r\Delta t}
 E[\mc{O}_{i}|S_{i-1}]\right\} ,
 \eeq
where $E[\mc{O}_{i}|S_{i-1}]$ is the conditional
expectation value of $\mc{O}_{i}$, i.e. its expectation value under 
the hypothesis of having  the price $S_{i-1}$ at the time
$t_{i-1}$. In this way, to get the actual price $\mc{O}_{0}$, 
it is necessary to proceed backward
in time and calculate $\mc{O}_{n-1},\ldots,\mc{O}_{1}$, where the value
$\mc{O}_{n}$ of the option at maturity is nothing but
$\mc{O}_{n}^{Y}(S_{n})$. It is therefore clear that evaluating the price
of an option with early exercise features means to simulate the evolution of the
underlying asset price (to obtain the $\mc{O}_{i}^{Y}$) and to
calculate a (usually large) number of expectation
conditional probabilities.\\

\subsection{Standard numerical procedures}
\label{optionn}

To value derivatives when analytical formulae are not available, 
appropriate numerical techniques have to be advocated. They involve 
the use of Monte Carlo (MC) simulation, binomial trees (and their improvements)
and finite difference methods~\cite{Hu,WDH}.

A natural way to simulate price paths is to discretize
eq.~(\ref{eq:itow}) as
 \[
 \ln S(t+\Delta t)-\ln S(t)= A \Delta t
 + \sigma \epsilon \sqrt{\Delta t} ,
 \]
 or, equivalently,
 \beq S(t+\Delta
 t)=S(t)\exp{\left[ A \Delta t +
 \sigma \epsilon \sqrt{\Delta t}\right]},
 \label{eq:sim}
 \eeq
which is correct for any $\Delta t >0$, even if finite.
Given the spot price $S_{0}$, i.e. the
price of the asset at time $t=0$, one can extract from a standardized 
normal distribution a
value $\epsilon_{k},k=1,\ldots,n$ for the random variable $\epsilon$ to 
simulate one
possible path followed by the price by means of eq.~(\ref{eq:sim}):
 \[
 S(k\Delta t)=S((k-1)\Delta t)\exp{\left[
 A \Delta t + \sigma \epsilon_{k}
 \sqrt{\Delta t}\right]}.
 \]
Iterating the procedure $m$ times, one can simulate 
$m$ price paths $\{ (S_{0},S_{1}^{(j)},
S_{2}^{(j)},\\ \ldots,S_{n}^{(j)}\equiv S_{T}^{(j)}):j=1,\ldots,m \}$,
to which apply the procedure exemplified in 
Section \ref{optiont} and evaluate the price of the option. In such a MC 
simulation of the stochastic dynamics of asset price (Monte Carlo random 
walk) the mean values
$E[\mc{O}_{i}|S_{i-1}],i=1,\ldots,n$ can be simply obtained as
 \[
 E[\mc{O}_{i}|S_{i-1}]=\frac{\mc{O}_{i}^{(1)}+\mc{O}_{i}^{(2)}
 +\cdots+\mc{O}_{i}^{(m)} }{m} ,
 \]
with no need to calculate transition probabilities because, through
the extraction of the possible $\epsilon$ values, the paths 
are automatically weighted according to the
probability distribution function of eq.~(\ref{eq:pdf}).\\
Unfortunately, this method leads to an estimated value whose 
numerical error
is proportional to $m^{-1/2}$. Thus, even if it is powerful 
because of the possibility to control the paths and to
impose additional  constrains (as it is usually required by 
exotic and path-dependent options), the MC random
walk is extremely time consuming when precise predictions are required 
and appropriate variance reduction procedures have to be 
used to save CPU time \cite{Hu}.\\
This difficulty can be overcome by means of the method of
the binomial trees and its extensions (see~\cite{Hu} and 
references therein), whose main idea 
stands in a deterministic choice of the possible paths to limit the number of
intermediate points. At each time step the price $S_{i}$ is
assumed to have only two choices: increase to the value
$uS_{i},u>1$ or decrease to $dS_{i},0<d<1$, where the parameters 
$u$ and $d$ are
given in terms of $\sigma$ and $\Delta t$ in such a way to give the 
correct values for the mean and variance of stock price changes over 
the time interval $\Delta t$.
Also finite difference methods are known in the literature~\cite{Hu} as an
alternative to time-consuming MC simulations. They provide 
the value of the derivative by solving the differential equation satisfied
by the derivative, by converting it into a difference equation. Although
tree approaches and finite difference methods are known to be faster than 
the MC random walk, they are difficult to apply when a detailed
control of the history of the derivative is required and are also
computationally time consuming when a number of stochastic variables 
is involved~\cite{Hu}. It follows that the development of efficient and fast 
computational algorithms to price financial derivatives is still 
a key issue in financial analysis.

\section{The path integral method} 
\label{path}

The path integral method is an integral formulation of the 
dynamics of a stochastic process. It is a suitable framework 
for the calculation of the transition probabilities associated 
to a given stochastic process, which is seen as the convolution of
an infinite sequence of infinitesimal short-time steps~\cite{Rt,S}.\\ 
For the problem of option pricing, 
the path integral method can be employed for the 
explicit calculation of the expectation values 
of the quantities of financial interest, given by 
integrals of the form~\cite{Rt}
 \beq
 E[\mc{O}_{i}|S_{i-1}]=\int dz_{i}
 p(z_{i}|z_{i-1})\mc{O}_{i}(e^{z_{i}}) ,
 \label{eq:mi}
 \eeq
where $z=\ln S$ and $p(z_{i}|z_{i-1})$ is the transition probability.
$E[\mc{O}_{i}|S_{i-1}]$ is the conditional expectation value of some functional $\mc{O}_{i}$
of the stochastic process. For example, for an European call option at the maturity
$T$ the quantity of interest will be ${\rm max}\,\{S_T-X,0\}$, $X$ being the strike price.
As already emphasized, and discussed in the 
literature~\cite{Hu,WDH,bpa,Rt2,M}, 
the computational complexity associated 
to this calculation is generally great: in the case 
of exotic options, with path-dependent and early exercise features, 
integrals of the type (\ref{eq:mi}) can not be analytically solved. As a
consequence,  we demand two things from a path integral framework:
a very quick way to estimate the transition probability 
associated to a stochastic process (\ref{eq:itow}) and a
clever choice of the integration points with which evaluate the
integrals (\ref{eq:mi}). In particular, our aim is to 
develop an efficient calculation of the probability distribution 
without losing information on the path followed by the asset price
during its time evolution.

\subsection{Transition probability}
\label{tp}
The probability distribution function related to a SDE verifies the so-called
Chapman-Kolmogorov equation~\cite{PB}, i.e. the relation
 \beq
 p(z''|z')=\int dz p(z''|z)p(z|z') ,
 \label{eq:ck}
 \eeq
which states that the  probability (density) of a transition
from the value $z'$ (at time $t'$) to the value $z''$ (at time
$t''$) is the ``summation'' over all the possible
intermediate values $z$ of the probability of separate and
consequent  transitions $z'\to z$, $z\to z''$.\\
As a consequence, if we consider a finite time interval $[t',t'']$ and we
apply a time slicing, 
by considering $n+1$ subintervals of length  $\Delta t\doteq
(t''-t')/n+1$, we can write, by iteration of eq.~(\ref{eq:ck})
 \[
 p(z''|z')=\int_{-\infty}^{+\infty}\!\!\!\cdots\int_{-\infty}^{+\infty}\!\!\!\!dz_{1}\cdots
 dz_{n}p(z''|z_{n})p(z_{n}|z_{n-1})\cdots p(z_{1}|z') ,
 \]
which, thanks to eq.(\ref{eq:gauss}), can be written as
 \beq
\int_{-\infty}^{+\infty}\!\!\!\cdots\int_{-\infty}^{+\infty}\!\!\!\!dz_{1}\cdots
 dz_{n} \frac{1}{\sqrt{(2\pi \sigma^{2}\Delta t)^{n+1}}}
 \cdot\exp{\left\{-\frac{1}{2\sigma^{2}\Delta t}\sum_{k=1}^{n+1}\left[
 z_{k}-(z_{k-1}+A\Delta t)\right]^{2}\right\}}.
 \label{eq:pd}
 \eeq
In the limit  $n\to\infty$, $\Delta t\to 0$ such that $(n+1)\Delta t=
(t''-t')$ (infinite sequence of infinitesimal time steps), 
the expression (\ref{eq:pd}), as explicitly shown in Ref.~\cite{Rt},
 exhibits a Lagrangian structure and it is possible to
express the transition probability in the path integral formalism as
a convolution of the form~\cite{Rt}
 \[
 p(z'',t''|z',t')=\int_{\mc{C}}\mc{D}[\sigma^{-1}\tilde{z}]
 \exp{\left\{-\int_{t'}^{t''}L (\tilde{z}(\tau),\dot{\tilde{z}}(\tau);\tau)
 d\tau\right\}} ,
 \]
 where $L$ is the Lagrangian
 \[
 L (\tilde{z}(\tau),\dot{\tilde{z}}(\tau);\tau)=\frac{1}{2\sigma^{2}}\left[
 \dot{\tilde{z}}(\tau)-A\right]^{2} ,
 \]
 and the integral is performed (with functional measure $\mc{D}[\cdot]$)
  over the paths
 $\tilde{z}(\cdot)$
 belonging to $\mc{C}$, i.e. all the continuous functions with constrains
 $\tilde{z}(t')\equiv z'$, $\tilde{z}(t'')\equiv z''$.
 As carefully discussed 
 in Ref.~\cite{Rt}, a path integral is well defined only if both a continuous formal
 expression and a discretization rule are given. As done in many applications, 
 the It\^o prescription is adopted in the present work.\\
A first, na\"{\i}ve evaluation of the transition probability (\ref{eq:pd}) 
can be performed via Monte Carlo simulation, by writing 
eq.~(\ref{eq:pd}) as
 \beq
 p(z'',t''|z',t')=\int_{-\infty}^{+\infty}\cdots\int_{-\infty}^{+\infty}\prod_{i}^{n}dg_{i}
\frac{1}{\sqrt{2\pi \sigma^{2}\Delta
t}}\exp{\left\{-\frac{1}{2\sigma^{2}\Delta t} \left[ z''-(z_{n}+A\Delta
t)\right]^{2}\right\}} ,
 \eeq
in terms of the variables $g_{i}$ defined by the relation
 \beq
 dg_{k}\doteq
 \frac{dz_{k}}{\sqrt{2\pi \sigma^{2} \Delta t
 }}\exp{\left\{-\frac{1}{2\sigma^{2}\Delta t} \left[
 z_{k}-(z_{k-1}+A\Delta t)\right]^{2}\right\}} ,
 \eeq
 and extracting each $g_{i}$ from a gaussian distribution of mean
$z_{k-1}+A\Delta t$ and variance  $\sigma^{2}\Delta t$. However, 
as we will see, this method requires a large number of calls 
to obtain a good precision. This is due to the fact that each $g_{i}$
is related to the previous $g_{i-1}$, so that 
this implementation of the path integral approach can be seen 
to be equivalent to a na\"{\i}ve MC simulation of random walks, 
with no variance reduction. 

By means of appropriate
manipulations~\cite{S} of the integrand entering eq.~(\ref{eq:pd}),
it is possible, as shown in the following, to 
obtain a path integral expression which
will contain a factorized integral with a constant kernel and a
consequent variance reduction. We will refer to this second 
implementation of the method as path integral with importance sampling.\\
If we define $z'' \doteq z_{n+1}$ 
and $y_{k}\doteq z_{k}-kA\Delta t$, $k=1,\ldots,n$, we
can express  the transition probability distribution as
 \beq
\int_{-\infty}^{+\infty}\cdots\int_{-\infty}^{+\infty}dy_{1}\cdots
 dy_{n} \frac{1}{\sqrt{(2\pi \sigma^{2}\Delta t)^{n+1}}}
 \cdot\exp{\left\{-\frac{1}{2\sigma^{2}\Delta t}\sum_{k=1}^{n+1}
 [y_{k}-y_{k-1}]^{2}\right\}},
 \label{eq:qq}
 \eeq
 in order to get rid of the contribution of the drift parameter.
Now let us extract from the argument of the exponential function a
quadratic form
 \[
 \sum_{k=1}^{n+1}[y_{k}-y_{k-1}]^{2}=y_{0}^{2}-2y_{1}y_{0}+y_{1}^{2}+y_{1}^{2}-2y_{1}
 y_{2}+\ldots+y_{n+1}^{2}=
 \]
 \beq =y^{t}M y+[y_{0}^{2}-2y_{1}y_{0}+y_{n+1}^{2}-2y_{n}y_{n+1}] ,
 \label{eq:qfm}
 \eeq
by introducing the $n$-dimensional array $y$ and the 
$n\mathrm{x}n$ matrix
$M$ defined as \beq
 y=\left(
 \begin{array}{l} y_{1}\\y_{2}\\ \vdots \\ \vdots \\ y_{n}
 \end{array} \right), 
 \qquad \qquad \qquad \qquad
 M=\left( \begin{array}{llllll}
  2&-1&0&\cdots&\cdots&0 \\
  -1&\ 2&-1&0&\cdots&0 \\
  0&-1&\ 2&-1&\cdots&0\\
  0&\cdots&-1&\ 2&-1&0 \\
  0&\cdots&\cdots&-1&\ 2&-1\\
  0&\cdots&\cdots&\cdots&-1&\ 2 \end{array} \right),
 \eeq
where $M$ is a real, symmetric, non singular and tridiagonal matrix. 
In terms of the eigenvalues $m_{i}$ of the matrix $M$, 
the  
contribution in eq.~(\ref{eq:qfm}) can be written as
 \beq
 y^{t}My=w^{t} O^{t} MO w=w^{t}M_{d}w=\sum_{i=1}^{n}m_{i}w_{i}^{2},
 \eeq
by introducing the orthogonal matrix $O$ which diagonalizes
$M$, with $w_i = O_{ij} y_j$.
Because of the orthogonality of $O$, the Jacobian
\[
J=\mathrm{det}\left|\derr{w_{i}}{y_{k}}\right|=\mathrm{det}|O_{ki}| ,
\]
of the transformation $y_{k}\to w_{k}$ equals 1, so that
$\prod_{i=1}^{n} dw_{i}=\prod_{i=1}^{n} dy_{i}$. 
Thanks to eqs.~(16)-(17), and after some algebra, 
eq.~(\ref{eq:qfm}) can be written as 
 \[
 \sum_{k=1}^{n+1}[y_{k}-y_{k-1}]^{2}=\sum_{i=1}^{n}m_{i}w_{i}^{2}+y_{0}^{2}-2y_{1}y_{0}
 +y_{n+1}^{2}-2y_{n}y_{n+1}=
 \]
 \beq
 \sum_{i=1}^{n}m_{i}\left[w_{i}-\frac{(y_{0}O_{1i}+y_{n+1}O_{ni})}{m_{i}}\right]^{2}
 +y_{0}^{2}+y_{n+1}^{2}-
 \sum_{i=1}^{n}\frac{(y_{0}O_{1i}+y_{n+1}O_{ni})^{2}}{m_{i}}.
 \eeq
Now,  if we introduce new
variables $h_{i}$ obeying the relation
 \beq
 dh_{i}\doteq\sqrt{ \frac{m_{i}}{2\pi \sigma^{2}\Delta t}}
 \exp{ \left\{ -\frac{m_{i}}{2 \sigma^{2}\Delta t}
 \left[w_{i}-\frac{(y_{0}O_{1i}+y_{n+1}O_{ni})}
 {m_{i}}\right]^{2}\right\}}dw_{i},
 \eeq
it is possible to express the finite-time 
probability distribution $p(z''|z')$ as
 \[
 \int_{-\infty}^{+\infty}\cdots\int_{-\infty}^{+\infty}\prod_{i=1}^{n}dy_{i}
 \frac{1}{\sqrt{(2\pi \sigma^{2}\Delta t)^{n+1}}}
 \exp{\left\{-\frac{1}{2\sigma^{2}\Delta t}\sum_{k=1}^{n+1}
 [y_{k}-y_{k-1}]^{2}\right\}}=
 \]
 \[
 =\int_{-\infty}^{+\infty}\cdots\int_{-\infty}^{+\infty}\prod_{i=1}^{n}dw_{i}
 \frac{1}{\sqrt{(2\pi \sigma^{2}\Delta t)^{n+1}}}
 e^{-(y_{0}^{2}+y_{n+1}^{2})/2\sigma^{2}\Delta t}\cdot
 \]
 \[
 \cdot
 \exp{\left\{-\frac{1}{2\sigma^{2}\Delta
 t}\sum_{i=1}^{n}\left[m_{i}\left(w_{i}-\frac{(y_{0}O_{1i}+y_{n+1}O_{ni})}{m_{i}}\right)^{2}-\frac{(y_{0}O_{1i}+y_{n+1}O_{ni})^{2}}{m_{i}}
 \right]\right\}}=
 \]
 \beq
 \!\int_{-\infty}^{+\infty}\!\!\cdots\!\!\int_{-\infty}^{+\infty}\prod_{i=1}^{n}dh_{i}
 \frac{1}{\sqrt{2\pi \sigma^{2}\Delta t \mathrm{det}(M)}}\cdot
 \exp{ \left \{ -\frac{1}{2\sigma^{2}\Delta t} \left[
 y_{0}^{2}+y_{n+1}^{2}+ \sum_{i=1}^{n}
 \frac{(y_{0}O_{1i}+y_{n+1}O_{ni})^{2}}{m_{i}} \right ] \right \}} .
 \label{eq:main}
 \eeq
 Equation~(\ref{eq:main}) is one of the main results of the present work.
  Actually, the probability distribution function,
   as given by eq.~(\ref{eq:main}), is 
  an integral whose kernel is a constant function
(with respect to the integration variables) and which can be
factorized into the $n$ integrals
 \beq
 \int_{-\infty}^{+\infty} dh_{i} \exp{ \left \{
 -\frac{1}{2\sigma^{2}\Delta t}
 \frac{(y_{0}O_{1i}+y_{n+1}O_{ni})^{2}}{m_{i}} \right \} },
 \eeq
given in terms of the $h_{i}$, which are gaussian variables 
that can be extracted 
from a normal distribution
with mean $(y_{0}O_{1i}+y_{n+1}O_{ni})^{2}/m_{i}$ and variance
$\sigma^{2}\Delta t/m_{i}$. Differently to the first, na\"{\i}ve 
implementation of the path integral, now each $h_{i}$ is no longer
dependent on the previous $h_{i-1}$, and importance sampling over 
the paths is automatically accounted for. \\
It is worth noticing that, by means of the extraction of 
the random variables $h_{i}$, we are creating price 
paths, since at each intermediate time $t_{i}$ the asset price is given 
by
 \begin{equation}
 S_{i}=\exp{\{\sum_{k=1}^{n} O_{ik}h_{k}+iA\Delta t\}}.
 \label{eq:ass}
 \end{equation}
Therefore, the path integral algorithm can be easily adapted to the cases 
in which the derivative to be valued has, in the time
interval $[0,T]$, additional
constraints, as in the case of interesting path-dependent options, 
such as Asian and barrier options~\cite{Hu}.\\
The results of the 
two realizations of the path integral method here discussed 
will be compared in 
Section~\ref{numerical}.

\subsection{Integration points}
\label{pathi}

Thanks to the method illustrated in Section~\ref{tp}, a powerful
and fast tool to compute the transition probability 
in the path integral framework is available and it can be employed if
we need to value a generic option with maturity $T$ and with
possibility of anticipated exercise at times $t_{i}=i\Delta t$
($n\Delta t\doteq T)$. As a consequence of this time slicing, 
one must numerically evaluate
 $n-1$ mean values of the type (9), 
 in order to
 check at any time $t_i$, and for any value of the stock price, 
 whether early exercise is more convenient with respect to
 holding the option for a future time. To keep under
control the computational complexity and the time of execution, 
it is mandatory to 
limit as far as possible the number of points for the integral
evaluation. This means that we would like to have a linear growth
of the number of integration points with the time.\\
Let us suppose to evaluate each mean value
 \[
 E[\mc{O}_{i}|S_{i-1}]=\int dz_{i}
 \, p(z_{i}|z_{i-1})\mc{O}_{i}(e^{z_{i}}) ,
 \]
with $p$  integration points, i.e. considering only $p$ fixed
values for $z_{i}$. To this end, we can create a grid of
possible prices, according to the dynamics of the 
stochastic 
process as given by eq.~(\ref{eq:itow})
 \beq
 z(t+\Delta t)-z(t)=\ln S(t+\Delta t)-\ln S(t)=A\Delta t+\epsilon
 \sigma\sqrt{\Delta t}.
 \eeq
Starting from $z_{0}$, we thus evaluate the expectation value
$E[\mc{O}_{1}|S_{0}]$ with  $p=2m+1,m\in\mathbb{N}$ values of
$z_{1}$ centered\footnote{Let us recall  that between two possible
exercise times the probability distribution function
 is gaussian and it is therefore symmetrical with
respect to its mean value.} on the mean value
$E[z_{1}]=z_{0}+A\Delta t$ and which differ from each other of a
quantity of the order of $\sigma\sqrt{\Delta t}$
 \[
 z_{1}^{j}\doteq z_{0}+A\Delta
 t+j\sigma\sqrt{\Delta t}, \quad \quad \ j=-m,\ldots,+m.
 \]
Going on like this, we can evaluate each expectation value
$E[\mc{O}_{2}|z_{1}^{j}]$ obtained from each one of the $z_{1}$'s
created above with $p$ values for $z_{2}$ centered around the mean
value
 \[
 E[z_{2}|z_{1}^{j}]=z_{1}^{j}+A\Delta t=z_{0}+2A\Delta
 t+j\sigma\sqrt{\Delta t}.
 \]
 
Iterating the procedure until the maturity, we create a 
deterministic grid of points
such that, at a given time $t_{i}$, there are $(p-1)i +1$ values of
$z_{i}$, in agreement with the request of linear growth.\\
This procedure of selection of integration points, together with 
the calculation of the transition 
probability previously described, is the basis of our 
path integral simulation of the 
price of a generic option.

 \begin{figure}
 \begin{center}
 \vspace{3mm}
 \includegraphics[scale=0.5]{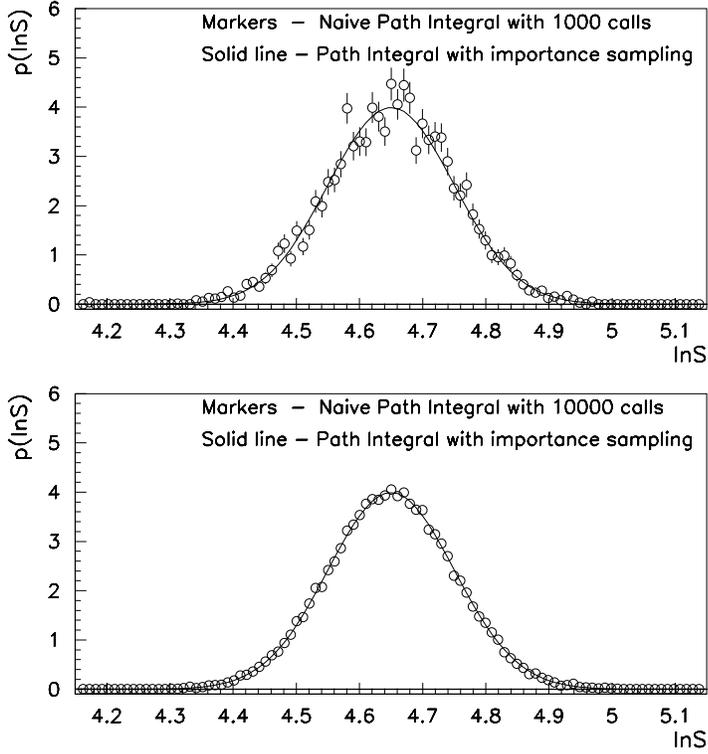}
 \caption{Simulation of the transition probability distribution 
 in the BSM model as a function of the logarithm of stock prices
 via the two path integral methods discussed in the text: the na\"{\i}ve path
 integral implementation for $10^3$ and $10^4$ Monte Carlo calls 
 (markers) is compared with the path integral implementation with 
 importance sampling (solid line).}
 \label{fig:rho}
 \end{center}
 \end{figure}

\section{Numerical results and discussion}
\label{numerical}

By applying the results derived in Section~\ref{path}, we have at 
disposal an efficient path integral algorithm both for 
the calculation of transition probabilities and the evaluation of option
prices. In the present section, the application of the method
to European and American options in the BSM model 
is illustrated and comparisons with the results obtained with 
the standard procedures known in the literature are shown.

First, the path integral simulation of the probability distribution 
of the logarithm of the stock prices, $p(ln S)$, as a function of the 
logarithm of the stock price, 
for a BSM-like stochastic model, as given by eq.~(\ref{eq:ito}), is shown in 
Fig.~\ref{fig:rho}. The parameters used in the simulation are: $S_0 = 100$,
$X = 110$, $\mu = 0.05$, $\sigma = 0.1$, $t=0$ year and $T = 1$~year, 
with 100 time slices. As can be seen, the expected lognormal distribution of 
the stock prices is correctly reproduced by the path integral 
numerical simulation. The plot shows 
a comparison of the calculation of $p(ln S)$ as obtained by means of the 
two path integral algorithms described 
in Section~\ref{tp}. The markers correspond 
to the na\"{\i}ve path integral computation of the probability distribution, 
without variance 
reduction, for $10^3$ (upper plot) and $10^4$ (lower plot) 
MC iterations. The error bars indicate the $1\sigma$ statistical error
of the MC calculation. The solid line is the prediction for $p(ln S)$ as 
obtained with the path integral simulation with importance sampling. In 
such a case, only two calls for each variables 
$h_i$ are needed to correctly fit a gaussian 
distribution, the numerical error being totally negligible and the 
algorithm very fast, with a typical execution time of a few seconds 
on a PentiumIII 500 Mhz PC. On the contrary, 
the first path integral implementation
is much less accurate and CPU time consuming. This is a consequence of the 
fact that, in the path integral simulation with importance sampling, 
the presence of constant integration kernel squeezes to zero the
standard estimation error. The diagonalization of the tridiagonal 
matrix $M$, which is a basic ingredient of the efficient path integral
algorithm developed, is performed according to the standard numerical 
procedure described in Ref.~\cite{nr}, realized by means of the routine 
F02FAF of the NAG program library~\cite{nag}, while the generation of 
the gaussian
variables $h_i$ follows from the routine RNORML of the CERN program library.\\
\begin{table}
 \caption{Price of an European put option in the BSM 
 model for the parameters $t=0$ year, $T=0.5$ year, $r=0.1$,
 $\sigma=0.4$, $X = 10$,  as a function of different stock prices $S_0$. 
 100 time slices are used in the path integral simulation.}
 \begin{tabular}{c c c c c}
 \hline
 \hline $S_{0}$ &  analytical & binomial tree & GFDNM & path integral\\
 \hline
 \hline
 6.0 & 3.558 & 3.557 & 3.557 & 3.558\\
 8.0 & 1.918 & 1.917 & 1.917 &  1.918\\
 10.0 & 0.870 & 0.866 & 0.871 &  0.870\\
 12.0 & 0.348 & 0.351 & 0.349 &  0.348\\
 14.0 & 0.128 & 0.128 & 0.129 &  0.128\\
 \hline
 \hline
 \end{tabular}
 \end{table}
As previously emphasized, the knowledge of the asset price $S_{i}$ 
at each intermediate time $t_{i}$ through eq.~(\ref{eq:ass}) gives the 
possibility of applying the procedure of calculation of the 
transition probability illustrated in Section \ref{tp}
also to those types of financial derivatives whose payoff depends
on the price path of the underlying asset. For such exotic 
options, the fast computation of the transition probability is a 
basic ingredient of our path integral algorithm as an efficient 
alternative to time-consuming MC simulation.

Once the transition probability has been computed, the price of an option 
can be computed in a path integral approach as the conditional 
expectation value of a given functional of the stochastic process.
 For example, the price of an European call option will be given by
 \beq
  \mc{C} = e^{-r (T -t)} \int_{-\infty}^{+\infty} dz_f \, p(z_f,T|z_i,t) \,
  {\rm max} [e^{z_f} - X, 0],
 \eeq
while for an European put it will be
 \beq
  \mc{P} = e^{-r (T-t)} \int_{-\infty}^{+\infty} dz_f \, p(z_f,T|z_i,t) \,
  {\rm max} [X - e^{z_f}, 0] ,
 \eeq
 where $r$ is the risk-free interest rate. Therefore just one-dimensional integrals need to be evaluated. They 
can be precisely computed with standard quadrature rules.  
In our calculation, the one-dimensional integrals are simply 
performed with a standard 
trapezoidal rule, cross-checked with the routine of adaptive integration 
D01EAF from the NAG library~\cite{nag}. A sample of the results 
obtained for an European put option in the BSM model is shown in Tab.~1. 
The predictions of our approach, indicated 
as path integral, are compared with results available in the literature, as 
quoted in Ref.~\cite{Rt2}. In Tab.~1, the entries 
correspond to the analytical results, 
the results by binomial trees, and the results of the Green Function 
Deterministic Numerical Method (GFNDM) developed in Ref.~\cite{Rt2}.
As can be noticed, our results are in perfect agreement with the
analytical predictions, while 
the differences with the other numerical procedures
are within the 1\% level. The errors in our numbers as due to numerical integration 
are not specified being well below the digits quoted. 
\begin{figure}
 \begin{center}
 \vspace{3mm}
 \includegraphics[scale=0.5]{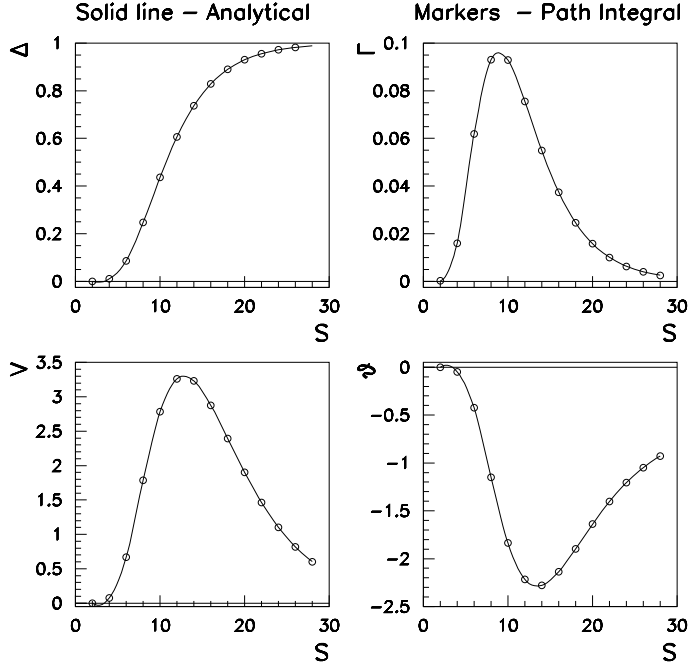}
 \caption{The Greek letters delta ($\Delta$), 
 gamma ($\Gamma$), vega ($V$) and theta ($\vartheta$) for an European 
 call option in the BSM model, as functions of the stock price. 
 The lines are the 
 analytical predictions, while the markers are the results of 
 our path integral algorithm. }
 \label{fig:greeks}
 \end{center}
 \end{figure}

Further numerical results of our path integral method are shown 
in Fig.~\ref{fig:greeks}, 
where the behaviour of the Greek letters~\cite{Hu} of an European call option 
in the BSM model is plotted as 
a function of the stock price. The solid lines correspond to the analytical
predictions, while the markers are the results by 
our path integral approach. The model parameters 
used in the simulation are: $X = 10$, $r = 0.1$, $\sigma = 0.4$, 
$t = 0$~year and $T = 0.5$~year. The four Greeks shown are 
important variables to manage the risk associated to an option 
and are defined as~\cite{Hu}
\begin{eqnarray*}
\Delta = \frac{\partial \mc{C}}{\partial S}, \qquad \quad
\Gamma = \frac{\partial^2 \mc{C}}{\partial S^2}, 
\qquad \quad V = {\rm Vega} = \frac{\partial \mc{C}}{\partial \sigma}, \qquad \quad
\vartheta = \frac{\partial \mc{C}}{\partial t} , \\
\end{eqnarray*}
where $\mc{C}$ is the price of the European call option. 
They are computed through 
standard numerical differentiation of the numerical value 
$\mc{C}$
returned by the path integral, by means of the 
NAG routine D04AAF~\cite{nag}. Also for the Greeks, there is perfect agreement
between the results obtained via our path integral 
simulation and the analytical predictions.

To test the reliability of the sampling 
over the integration points discussed in Section \ref{pathi}, we
present results for the particular case of the price of an American option 
in the BSM model in Tab.~2, as obtained with 
the grid technique described in Section \ref{pathi}. Comparisons 
with independent results available in the literature are also 
shown. As can be seen from Tab.~2, there is generally a good agreement
of our path integral results with those known in the 
literature~\cite{Rt2} and obtained by means of the binomial tree, of the finite difference
method and of the GFDNM method. It is worth noticing that our results in the 
path integral framework require only a few seconds on
a PentiumIII 500MhZ PC.

\begin{table}
 \caption{Price of an American put option in the BSM 
 model for the parameters $t=0$ year, $T=0.5$ year, $r=0.1$,
 $\sigma=0.4$, $X = 10$, as a function different stock prices $S_0$. 
 The path integral simulation is
 performed with  200 time slicings and $p=13$ integration points.}
 \begin{tabular}{c c c c c}
 \hline
 \hline $S_{0}$ &  finite difference & binomial tree & GFDNM & path integral \\
 \hline
 \hline
 6.0 & 4.00 & 4.00 & 4.00 & 4.00\\
 8.0 & 2.095 & 2.096 & 2.093 &  2.095\\
 10.0 & 0.921 & 0.920 & 0.922 &  0.922\\
 12.0 & 0.362 & 0.365 & 0.364 &  0.362\\
 14.0 & 0.132 & 0.133 & 0.133 &  0.132\\
 \hline
 \hline
 \end{tabular}
 \end{table}

\section{The limit of continuum and American options}
\label{american}

In the case of an American option, the possibility of exercise 
at any time up to the expiration date 
allows to develop, within the path integral formalism, 
a specific algorithm, which, as shown in the following, 
is precise and very quick.\\
Given the time slicing considered in Section~\ref{pathi}, the case of 
American options requires the limit $\Delta t\to 0$ which, putting 
$\sigma\to 0$,
 leads to a delta-like transition probability
 \[
 p(z,t+\Delta t|z_{t},t)\thickapprox\delta (z-z_{t}-A\Delta t) .
 \]
This means that, apart from volatility effects, the price $z_{i}$ 
 at time $t_i$ will have
a value remarkably close to the expected value $\bar{z}\doteq
z_{i-1}+A\Delta t$, given by the drift growth. Needless to say, 
if we should substitute the expression
$p(z_{i},i\Delta t|z_{i-1},(i-1)\Delta
t)\thickapprox\delta(z_{i}-\bar{z})$ inside the integrals (9), we would
neglect the role of the volatility and consider only a drift
growth of the asset prices. In order to take care of the 
volatility effects, a possible solution is to estimate the
integral of interest, i.e.
 \beq
 E[\mc{O}_{i}|S_{i-1}]=
 \int_{-\infty}^{+\infty}dz\,p(z|z_{i-1})\mc{O}_{i}(e^{z}) ,
 \label{eq:ri}
 \eeq
by inserting in eq.~(\ref{eq:ri}) the analytical expression for the 
$p(z|z_{i-1})$ transition probability
 \[
 p(z|z_{i-1})=\frac{1}{\sqrt{2\pi \Delta t \sigma^{2}}}
 \exp{\left\{-\frac{(z-z_{i-1}-A\Delta t)^{2}}{2\sigma^{2}\Delta
 t}\right\}}=
 \]
 \[
 \frac{1}{\sqrt{2\pi \Delta t \sigma^{2}}}
 \exp{\left\{-\frac{(z-\bar{z})^{2}}{2\sigma^{2}\Delta t}\right\}} ,
 \label{eq:pda}
 \]
together with a Taylor expansion of the kernel function
$\mc{O}_{i}(e^{z})\doteq f(z)$  around the expected value
$\bar{z}$. Hence, up to the second order in $z-\bar{z}$, the 
kernel function becomes
 \beq
 f(z)=f(\bar{z})+(z-\bar{z})f'(\bar{z})+\frac{1}{2}
 f''(\bar{z})(z-\bar{z})^{2}+O((z-\bar{z})^{3}) ,
 \eeq
which, together with eq.~(\ref{eq:pda}), yields
 \beq
 E[\mc{O}_{i}|S_{i-1}]=f(\bar{z})+\frac{\sigma^{2}}{2}f''(\bar{z}) ,
 +\ldots,
 \eeq
since the first derivative does not give contribution to eq.~(\ref{eq:ri}),
 being the integral of an odd function over the whole 
$z$ range. The second derivative can be numerically
estimated as
 \beq
 f''(\bar{z})=\frac{1}{\delta_{\sigma}^{2}}[f(\bar{z}+
 \delta_{\sigma})-2f(\bar{z})+f(\bar{z}-\delta_{\sigma})] ,
 \eeq
with $\delta_{\sigma}=O(\sigma\sqrt{\Delta t})$, as dictated by the 
dynamics of the stochastic process.\\
\begin{table}
 \caption{Price of an American put option with $t=0$~year, 
 $T=0.5$ year, $r=0.1$,
 $\sigma=0.4$, $X = 10$, as a function different stock prices $S_0$.
 The path integral results are obtained with 
 $\delta_{\sigma}=2\sigma\sqrt{\Delta t}$, 300 time slices and $p=3$.}
 \begin{tabular}{c c c c c}
 \hline
 \hline $S_{0}$ &  finite difference & binomial tree & GFDNM & path integral \\
 \hline
 \hline
 6.0 & 4.00 & 4.00 & 4.00 & 4.00\\
 8.0 & 2.095 & 2.096 & 2.093 &  2.095\\
 10.0 & 0.921 & 0.920 & 0.922 &  0.922\\
 12.0 & 0.362 & 0.365 & 0.364 &  0.362\\
 14.0 & 0.132 & 0.133 & 0.133 &  0.132\\
 \hline
 \hline
 \end{tabular}
 \end{table}  
It is worth noticing that each expectation value
$E[\mc{O}_{i}|S_{i-1}]$ can be now computed once 
$f(\bar{z})=\mc{O}_{i}(e^{z_{i-1}+A\Delta t})$ and $f(\bar{z}\pm
\delta_{\sigma})= \mc{O}_{i}(e^{z_{i-1}+A\Delta t\pm \delta_{\sigma}})$ 
are known. Consequently, if we employ the deterministic grid illustrated in
Section \ref{pathi}, it is enough to put $p=3$ to obtain reliable 
results, provided $\Delta t$ is taken sufficiently small. Actually, the 
results obtained with this simple semi-analytical procedure are 
shown in Tab.~3, using $\delta_{\sigma}=2\sigma\sqrt{\Delta t}$ for 
numerical differentiation and 300 time slices. For such path integral
evaluation of an American option the CPU time is negligible. The results 
are in nice agreement with those of other numerical procedures and in 
perfect agreement, as we explicitly checked for different model 
parameters, with those quoted in Tab.~2 as obtained 
with the path integral
algorithm discussed in Section \ref{pathi}.

\section{Concluding remarks and outlook}
\label{conclusion}

In this paper we have shown that the path integral approach 
to stochastic processes can be successfully applied to the 
problem of option pricing in financial analysis. In particular, 
an efficient implementation of the path integral method has 
been presented, in order to obtain fast and accurate 
predictions for a large class of financial derivatives, including
those with path-dependent and early exercise features. The key 
points of the algorithm are a careful evaluation of the 
transition probability associated to the stochastic model 
for the time evolution of the asset prices 
and a suitable choice of the integration points needed to evaluate 
the quantities of financial interest. Furthermore, a simple and 
very fast procedure to value American options has been derived, 
by exploiting the possibility of continuous exercise up to the 
expiration date.  

The results of the path integral algorithm have been carefully compared 
with those available in the literature for European and American options
in the BSM model and found to be in good agreement with the standard 
numerical procedures used in finance. The computational 
time of the algorithm developed is, to the best of our 
knowledge, competitive with the most efficient strategies 
used in finance. The method is 
general and it can be quite easily extended to cope 
with other financial derivatives (with path-dependent features) and other 
models beyond the dynamics of geometric Brownian motion.  

The natural developments of the path integral algorithm here presented 
concern the application of the method to value other 
kinds of quantities of financial interest, for which the analytical 
solution is not available or not accessible, and the extension of the 
method of option pricing to more realistic model of the financial dynamics, 
such as models with stochastic volatility~\cite{Hu,B,st} or beyond 
the BSM Gaussian limit~\cite{ms,bp,man,msn,mo,fmj},       
in order to
search for a better agreement with the real prices 
as observed in the real market. 
A further interesting perspective would be using the 
path integral algorithm as a benchmark to train neural networks. 

Such developments are by now under consideration.

\vskip 8pt\noindent
{\bf Acknowledgments} \\ 
We acknowledge collaboration of F.~Piccinini at the early stage
of this work. We wish to thank M.~Cersich, E.~Melchioni and A.~Pallavicini of 
FMR Consulting for useful discussions. We are also grateful to FMR 
Consulting for having provided the software for the analytical 
predictions of the price and Greeks of European options.


\begin{thebibliography}{20}
 \bibitem{BS} F. Black and M. Scholes, 
 Journal of Political Economics {\bf 72} (1973) 637.
 \bibitem{Me} R. Merton, Bell J. Econom. Managem. Sci. {\bf 4} (1973) 141.
 \bibitem{Hu} J.C. Hull, \emph{Options, Futures, and Other Derivatives},
 Fourth Edition, Prentice Hall, New Jersey, 2000.
 \bibitem{PB} W. Paul and J. Baschnagel, \emph{Stochastic Processes: from
 Physics to Finance}, Springer-Verlag, Berlin Heidelberg, 1999.
 \bibitem{WDH} P. Wilmott, J. Dewynne and S. Howinson, \emph{Option Pricing: 
 Mathematical Models and Computation}, Oxford Financial Press, Oxford, 1993.
 \bibitem{bpa} M.~Potters, J.-P.~Bouchaud and D.~Sestovic, Physica A 
 {\bf 289} (2001) 517.
 \bibitem{W} N. Wiener, Proc. Nat. Acad. Sci. (USA) {\bf 7}  (1952) 253; 
 Proc. Nat. Acad. Sci. (USA) {\bf 7} (1952) 294.
 \bibitem{K} M. Kac, Bull. Amer. Math. Soc. {\bf 72} (1966) 52.
 \bibitem{FH} R. P. Feynman, Rev. Mod. Phys. {\bf 20} (1948) 367;\\
 R.~P.~Feynman and A.~R.~Hibbs, \emph{Quantum Mechanics and 
 Path Integral}, McGraw-Hill, New York, 1965. 
 \bibitem{Rt}  E. Bennati, M. Rosa-Clot and S. Taddei, \emph{A Path Integral
 Approach to Derivative Security Pricing: I. Formalism and Analytical Results},
 Int. Journ. Theor. Appl. Finance {\bf 2} (1999) 381, 
 {\tt cond-mat/9901277}.
 \bibitem{Rt2} M. Rosa-Clot and S.Taddei, \emph{A Path Integral
 Approach to Derivative Security Pricing: II. Numerical Methods}, 
 {\tt cond-mat/9901279}.
 \bibitem{B} B.E. Baaquie, J. Phys. I France {\bf 7} (1997) 1733.
 \bibitem{I} L. Ingber, \emph{High-Resolution Path-Integral Development of 
 Financial Options}, {\tt physics/0001048}.
 \bibitem{M} A. Matacz, \emph{Path Dependent Option Pricing: the Path Integral 
 Partial Averaging Method}, {\tt cond-mat/0005319}.
\bibitem{S} L.S.~Schulman, \emph{Techniques and Applications of Path Integration},
 John Wiley \& Sons, New York, 1981.
 \bibitem{nr} W.H. Press, B.P. Flannery, S.A. Teukolsky and W.T. Vetterling, 
 \emph{Numerical Recipes -  The Art of Scientific Computing}, Cambridge University
 Press, New York, 1989.
 \bibitem{nag} NAG Fortran library, The Numerical Algorithms Group Ltd, 
 Oxford.
 \bibitem{st} J.C.~Hull and A.~White, Journal of Finance {\bf 42} (1987) 281;\\
 J.~Masoliver and J.~Perello, \emph{A Correlated Stochastic Volatility Model 
 Measuring Leverage and Other Stylized Facts}, {\tt cond-mat/0111334}.
 \bibitem{ms} R.N. Mantegna and H.E. Stanley, \emph{An Introduction to Econophysics: Correlations
 and Complexity in Finance}, Cambridge University Press, Cambridge, 2000.
 \bibitem{bp} J.P. Bouchaud and M. Potters, 
 \emph{Theory of Financial 
 Risk: from Statistical Physics to Risk Management}, Cambridge University 
 Press, Cambridge, 2000.
 \bibitem{man} B.B. Mandelbrot, \emph{Fractals and Scaling in Finance}, 
 Springer-Verlag, Berlin Heidelberg, 1997.
 \bibitem{msn} R.N. Mantegna and H.E. Stanley, 
 Phys. Rev. Lett. {\bf 73} (1994) 2946; \\
 R.N. Mantegna and H.E. Stanley, Nature {\bf 383} (1996) 587;\\
 R.N. Mantegna, Z.~Pal\'agyi and H.E. Stanley, Physica A {\bf 274}
 (1999) 216.
 \bibitem{mo} A.~Matacz, \emph{Financial Modeling and Option Theory 
 with the Truncated Levy Process}, {\tt cond-mat/9710197}.
 \bibitem{fmj} Frederick Michael and M.D. Johnson, 
 \emph{Financial Market
 Dynamics}, {\tt cond-mat/0108017}.

\end{thebibliography}
\end{document}